\documentclass[aps,prl,superscriptaddress,12pt,a4paper]{revtex4-2}

\usepackage{graphicx}
\usepackage{dcolumn}
\usepackage{bm}
\usepackage{color}
\usepackage{tgbonum}

\usepackage{setspace} 
\begin{document}
\fontfamily{qag}\selectfont

\title{Emulsion Electrocoalescence in microfluidics: impact of local electric fields}

\author{David Van Assche}
\affiliation{Universit{\'e} de Bordeaux, CNRS, CRPP\unskip, UMR5031\unskip, 115 Avenue Albert Schweitzer\unskip, Pessac\unskip, 33600\unskip, France}

\author{Thomas Beneyton}
\affiliation{Universit{\'e} de Bordeaux, CNRS, CRPP\unskip, UMR5031\unskip, 115 Avenue Albert Schweitzer\unskip, Pessac\unskip, 33600\unskip, France}

\author{Alexandre Baron}
\affiliation{Universit{\'e} de Bordeaux, CNRS, CRPP\unskip, UMR5031\unskip, 115 Avenue Albert Schweitzer\unskip, Pessac\unskip, 33600\unskip, France}
\affiliation{Institut Universitaire de France, 75005 Paris, France}

\author{Jean-Christophe Baret}
\email{jean-christophe.baret@u-bordeaux.fr}
\affiliation{Universit{\'e} de Bordeaux, CNRS, CRPP\unskip, UMR5031\unskip, 115 Avenue Albert Schweitzer\unskip, Pessac\unskip, 33600\unskip, France}
\affiliation{Institut Universitaire de France, 75005 Paris, France}

\maketitle

\section*{abstract} 
\textbf{The mechanism of coalescence of aqueous droplet pairs under an electric field is quantitatively studied using microfluidics in quiescent conditions. We experimentally trap droplet pairs and apply electric fields with varying frequencies and formulation compositions. We find that the electrical resistance of the oil used as continuous phase controls the onset of electrocoalescence in quiescent conditions. We observe that the local field enhancement between droplets strongly depends on formulations but also on the number of droplets across the electrodes. These findings provide a better understanding of the onset of electrocoalescence and pave a route towards the rationalization of droplet-based microfluidics operations.}

\pagebreak
\clearpage

Electrocoalescence is the fusion of droplets dispersed in a continuous phase by the effect of electrostatic interactions. It is a well-established principle used in the petroleum industry for oil refining and separation from brine~\cite{Eow2001} and a physical mechanism proposed to explain heavy rains from atmospheric clouds~\cite{shavlov2023}. In the past two decades electrocoalescence  also found applications in the field of droplet-based microfluidics for the precise addition of reagents and for the controlled initiation of chemical reactions in droplets~\cite{Chabert2005, Link2006, Frenz2008, Niu2009, Abate2010}. This electrically switchable coalescence principle is important for the implementation of biological workflows that require automatized multi-step processing of material and reagents with external control~\cite{Autour2016,Sjostrom2013}. It completes the toolbox for active control of droplet actuation by electric field, together with high-throughput droplet sorting by dielectrophoresis~\cite{Fidalgo2008, Baret2009, Sciambi2015} and with the control of droplet production \cite{Malloggi2007, Tan2014, vanassche2023}. \\

Obviously, electrocoalescence requires an electric field to be applied between droplets. What is the critical value of the voltage to be applied at electrodes to obtain coalescence is a very basic question that remains unanswered. Microfluidics provides means to address this question with controlled emulsions and geometries. However, already in simple geometries, the literature indicates very diverse - and sometimes inconsistent - threshold values: Zagnoni \textit{et al.} obtain electrocoalescence of droplets in flow for applied fields of order 0.3 kV/m  \cite{Zagnoni2009}, Szymborski \textit{et al.} at 10 kV/m \cite{Szymborski2011}, Leary \textit{et al.} at 250 kV/m \cite{Leary2020}, Thiam \textit{et al.} at 1 MV/m \cite{Thiam2009}, Priest \textit{et al.} at 1.5 MV/m \cite{Priest2006}. The results might depend on the applied frequency \cite{Szymborski2011, Niu2009} or not \cite{Priest2006} and most certainly depend on the formulations used.\\

The theoretical description of electrocoalescence lies in the framework of electrohydrodynamics pioneered by Melcher and Taylor \cite{Melcher1969, taylor1966}. In brief, the problem has two sides: first, the applied electric field results in a net driving force on droplets: the force originates from the dipole-dipole interaction of the polarized droplets. The ratio of conductivities and of dielectric constants control both the droplet deformation and direction of the motion~\cite{Baygent1998, Dong2018}. For electrically conducting droplets immersed in a leaky dielectric the force is attractive in the direction of the field and repulsive in the orthogonal directions~\cite{atten1993, atten2006, Vlahovska2019, Zhang1995} bringing droplets together in line, not necessarily sufficient for full coalescence~\cite{Anand2020, Chen2023, Mhatre2015}. Second, the coalescence process, namely the destabilisation of the oil film separating the droplets is itself dependent on the electric field across the oil interface. In emulsions, droplets are stabilised against coalescence by surfactants. The (meta)-stability of the droplets is described by the interplay of interfacial forces and hydrodynamics, controlling the behaviour of colloidal systems (floculation, adhesion,\dots)~\cite{israelachvili2022}. In electric fields, the additional attractive interaction among droplets changes the stability of the film which ruptures for a sufficiently large surface charge at the droplet interface. The instability is described similar to the stability of a thin film in spinodal dewetting~\cite{Herminghaus1999}. The stability criterion of the film is expected to depend on formulations and as stated by Vlahovska~\cite{Vlahovska2019} `\textit{drops with more complex interfaces, e.g. coated with surfactants are likely to display additional rich electrohydrodynamics and merit further investigation.'}\\

Motivated by these identified unknowns and by the practical importance of the electrocoalescence principle in microfluidics, we experimentally analyze electrocoalescence. We use droplet-based microfluidics to produce monodisperse droplets, and destabilize droplet pairs in quiescent conditions. By varying the formulations of the emulsions, we find that the electrical resistivity of the oil phase – a parameter which is neglected in microfluidics – plays a significant role. The local field between the droplet pairs is found to be orders of magnitude larger than the applied field with an amplification controlled by the electric properties of the formulations. In addition, we observed experimentally that the field enhancement is further amplified when multiple droplets are confined between the electrodes: the threshold voltage for coalescence linearly decreases with the number of droplets, thereby explaining why emulsions appear to be experimentally more unstable to electric fields than droplet pairs. \\

\begin{figure}[!ht]
\centering
\includegraphics[width=\linewidth]{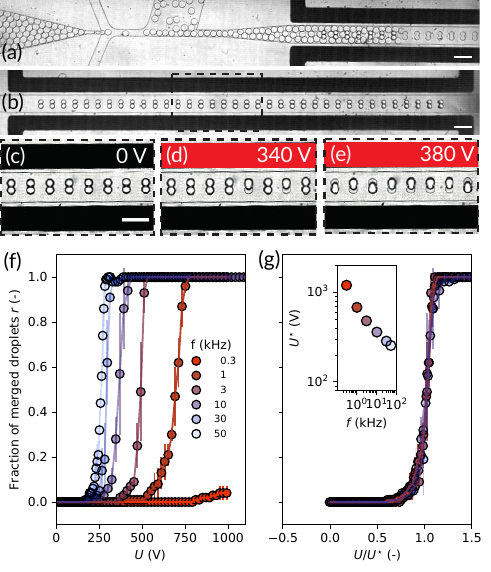}
\caption{Experimental workflow. (a) Micrograph of the coalescence chamber during droplet loading under a pressure control. (b) The coalescence chamber loaded with 36 droplet pairs. Droplets that were not trapped have been flushed out of the chamber. The electrodes are in black. (c-e) Zoom on the trapping array for $U=$ 0~V (c), 340~V (d) and 380~V (e) at $f=$ 10 kHz. (f) Fraction $r$ of coalesced droplet pairs as a function of the applied voltage $U$ for different frequencies $f$ of the electric field. The continuous phase is HFE 7500 with 5\%~w/w surfactant while the dispersed phase is millipore water. (g) Fraction $r$ of coalesced droplet pairs rescaled with the coalescence threshold voltage $U^{\star}$. Inset: coalescence threshold voltage $U^*$ as a function of the frequency $f$ of the electric field. ($N=$ 3). Scale bars are 200~$\mu$m.} 
\label{setup}
\end{figure}

We use microfluidics to study electrocoalescence in a controlled, quiescent environment. We couple a flow focusing junction~\cite{Anna2003} to an array of traps~\cite{Abbyad2011} designed to obtain 36 droplet pairs (\textcolor{red}{Fig.~\ref{setup}a-b}). Once the droplet pairs are trapped in the system, the flow is stopped. An electric field parallel to the droplet pair orientation is applied across electrodes patterned using the microsolidics methods~\cite{Siegel2007} (\textcolor{red}{Fig.~\ref{setup}c}, \textcolor{red}{Supp. Fig. S1, S2}). We first fix the formulation to a continuous phase of fluorocarbon oil HFE 7500 with 5\%~w/w of a non-ionic block-copolymer surfactant (Fluosurf, emulseo) and an aqueous phase of millipore water. The voltage across the electrodes is quasistatically increased. For a fixed field frequency of 10 kHz, at low voltages, typically below 100 V, all droplet pairs remain intact (\textcolor{red}{Fig.~\ref{setup}c}). At higher voltages, the fraction $r$ of coalesced drops, \textit{i.e.} the number of coalesced droplet pairs divided by the total number of droplet traps increases with $U$  (\textcolor{red}{Fig.~\ref{setup}d-e}, \textcolor{red}{Supp. Methods, Supp. Fig. S3}). $r$ shows a sharp increase from 0 to 1 in the range 200 V - 300 V and is a function of the applied frequency: the coalescence threshold voltage $U^{\star}$ defined as the voltage at which half of the droplet pairs coalesced is found to increase with decreasing frequency. For a given experimental condition, the threshold $U^{\star}$ is found to be consistent among repeats ($N$ = 3). $U^{\star}$ provides a means to rescale all data on a master curve (\textcolor{red}{Fig.~\ref{setup}g}), and is used as a quantitative measurement of the electrocoalescence efficiency for the different experimental conditions  (surfactant concentration, oil type, droplet ionic content, amount of droplets \dots). \\

\begin{figure*}[!ht]
\centering
\includegraphics*[width=\textwidth]{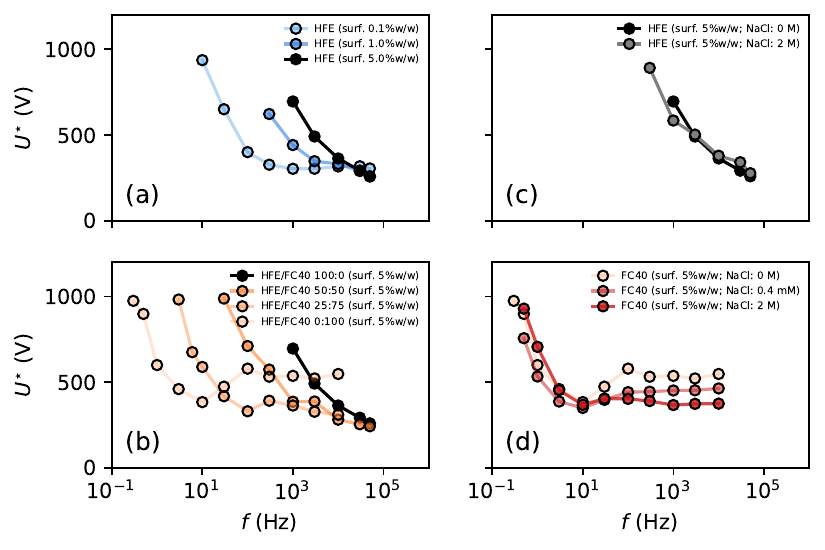}
\caption{Formulation dependence of electrocoalescence efficiency. (a) $U^{\star}$ is a function of the frequency of the electric field and of the surfactant concentration (for an oil phase of HFE 7500 and aqueous droplets of millipore water). (b) For a fixed surfactant concentration, $U^{\star}$ strongly depends on the nature of the oil (here the oil properties, especially the oil resistivity (Supp. Fig. S5) are varied using mixtures of FC 40 and HFE 7500). (c-d) The aqueous phase conductivity does not significantly influence the electrocoalescence as shown in the two extreme cases of HFE 7500 (c) and FC 40 (d) with varying salt concentrations ranging 0-2 M.}
\label{fig:experiments}
\end{figure*}

We first test the impact of formulations on electrocoalescence, by varying the surfactant concentration in the system when using millipore water as the aqueous phase and HFE 7500 as continuous phase. We perform the experiment with surfactant concentrations of 0.1\%~w/w, 1\%~w/w and 5\%~w/w. $r$ is measured for various frequencies of the electric field as a function of $U$, as above (\textcolor{red}{Supp. Fig. S3}), to obtain $U^{\star}$ as a function of the frequency $f$ (\textcolor{red}{Fig.~\ref{fig:experiments}a}). We confirm that for a fixed composition, $U^{\star}$ decreases with increasing $f$. But at low surfactant concentrations, we observe two regimes: a strong decrease of $U^{\star}$ at small frequencies $f<f_c$ and a plateau at large frequency $f>f_c$ where $f_c$ depends on the experimental conditions. These experimental results highlight the strong influence of surfactant concentration on the electrocoalescence efficiency. We could be tempted to interpret this effect as a stabilising role of the surfactant because the droplets are more stable against coalescence at high surfactant concentrations. Yet, it must be reminded that the interfacial coverage of the surfactant is almost unchanged at these concentrations \cite{Skhiri2012,Brosseau2014} which cannot explain such a wide variation in stabilisation efficiency (Supp. Methods, \textcolor{red}{Supp. Fig. S4}). We performed another set of experiments using mixtures of oils (HFE 7500 and FC 40) at fixed surfactant concentration of 5\%~w/w (\textcolor{red}{Fig.~\ref{fig:experiments}b}) which does not significantly change the interfacial tension and recover similar trends. 

\begin{figure*}[!h]
\centering
\includegraphics*[width=\linewidth]{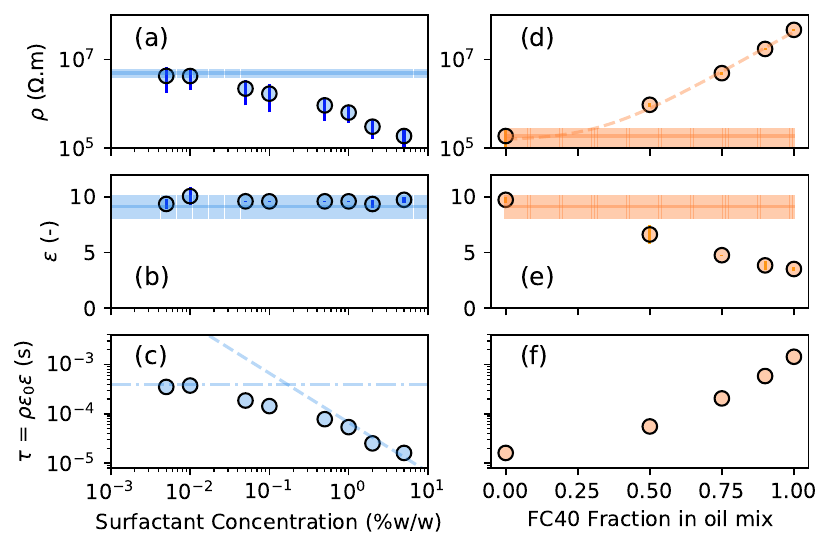}
\caption{Electrical properties of the continuous phase. (a-b) The electric resistivity and relative permittivity of HFE 7500 as a function of surfactant concentration. The solid and dotted lines represent the average and standard deviation of the measurement without surfactant. (c-d) The electric resistivity and relative permittivity of mixtures of FC 40 and HFE 7500 at a surfactant concentration of 5\% w/w. Error bars represent the standard deviation (N = 3).} \label{fig:S5}
\end{figure*}

The dependence of the electrical properties of the continuous phase must therefore be accounted for in the interpretation of our experimental results. The oil electrical resistivity $\rho$ is decreased by more than one order of magnitude with surfactant (\textcolor{red}{Fig.~\ref{fig:S5}}). For the sake of completeness, we also characterized the relative permittivity $\varepsilon$ of the oil phase which is shown to be independent on surfactant concentration (\textcolor{red}{Fig.~\ref{fig:S5}}). With these formulations, $\rho$ varies over two orders of magnitude with a moderate three-fold variation of $\varepsilon$ over the composition range (\textcolor{red}{Fig.~\ref{fig:S5}}).  

The dependence of the electrocoalescence efficiency therefore relates to the electric properties of the oil: we find that $U^{\star}(f)$ shifts towards lower frequencies when increasing the resistivity of the oil ($\rho$). Our experimental results confirm that the resistivity of the continuous phase is the main parameter controlling the electrocoalescence process. The case of FC 40 with 5\% w/w surfactant appears to be non-monotonous, with a minimum at intermediate frequencies ($f \sim$ 10 Hz). We repeated the experiments for the two limiting cases of FC 40 and HFE 7500 with 5\% w/w surfactant. We varied the electrical resistivity of the aqueous phase  $\hat{\rho}$ in the system by adding NaCl (0~M, $4\cdot10^{-4}$~M and 2~M) and repeated the experiments for HFE 7500  and FC 40 with 5\%~w/w surfactant (\textcolor{red}{Fig. \ref{fig:experiments}c-d}). We observe that: (i) the sharp dependence of $U^{\star}$ with $f$ at $f<f_c$ does not depend on $\hat{\rho}$, for both FC 40 and HFE 7500, (ii) the plateau value for $U^{\star}$ at $f>f_c$ decreases with decreasing $\hat{\rho}$. However this change in the plateau value of $U^{\star}$ is weak: a change of four orders of magnitude of the ion concentration changes $U^{\star}$ by less than 50\%. \\

With the geometry of the system, obtaining an analytical description of the field distribution is not possible. In order to gain insights on the underlying mechanisms, we numerically model our experimental setup using a minimal electric model in two dimensions (COMSOL, Supp. Methods, \textcolor{red}{Supp. Fig. S5-7}). We calculate the maximal value of the electric field in the domain comprising the droplets and oil phase. In the absence of droplets, the field is homogeneous and is a function of the frequency as expected from a simple voltage divider model. The dielectric constant of the PDMS coupled to the resisistivity of the oil determines the cutoff frequency of an equivalent high-pass filter (\textcolor{red}{Supp. Fig. S8}). We then define the droplets as non-deformable objects with a spacing distance of 100 nm or 1 micron between the edges of the droplets and compute the value of the field across the droplet-droplet interface for a pair of droplets. The actual spacing in the experiments is unknown but is expected to lie in this range of values for the formulations used here. The field across the oil film is significantly enhanced compared to the applied field, with a dependence on the thickness of the oil film and of the electric properties of the oil (\textcolor{red}{Supp. Fig. S6}). For a fixed gap size, the field increases at low frequency, shows a plateau and decrease at higher frequency. The cutoff frequencies are given by the PDMS/oil high-pass filter (at low frequency) as a first approximation and the water droplet charge relaxation at high frequency  $\hat{\tau} = \varepsilon_0 \hat{\varepsilon} \hat{\rho}$ (only dependent on the aqueous properties here). A simple electric model qualitatively reproduces the low frequency decay (\textcolor{red}{Supp. Fig. S8}). For the range of parameters tested experimentally $\hat{\tau} \lesssim$ 2 $\mu$s, which confirms that the electric properties of the aqueous phase marginally influences the electrocoalescence in the range of frequency tested here ($1/f >$ 10 $\mu$s). Hence, the droplet is approximated as a perfect conductor in all cases. We observe however that in the limit of high resistivity, the low frequency cutoff is not strictly given by the PDMS/oil RC circuit and deviations are observed, especially for the 100 nm gap (\textcolor{red}{Supp. Fig. S7}). \\

\begin{figure}[!ht]
\centering
\includegraphics*[width=\linewidth]{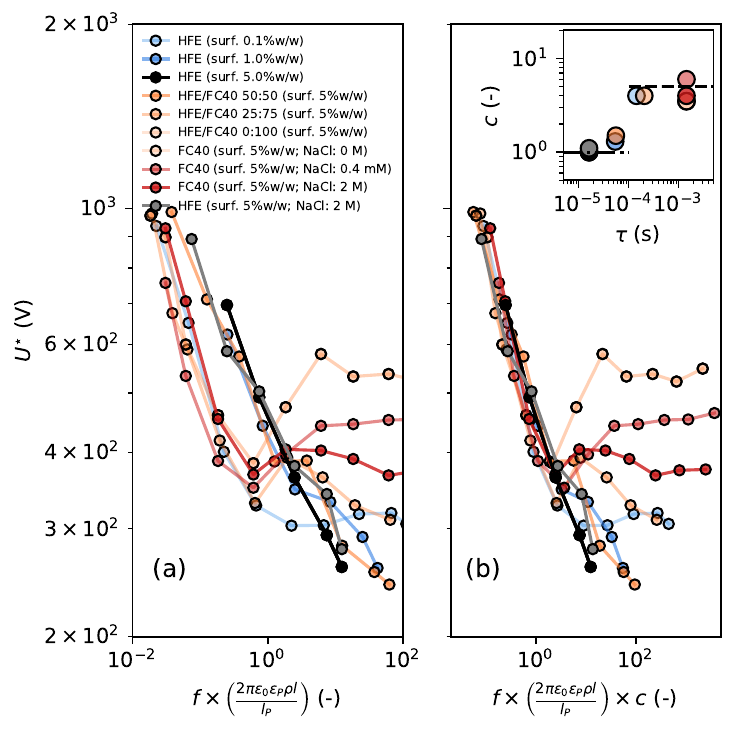}
\caption{Rescaling of the experimental data. (a) Experimental data rescaled with the PDMS/oil RC constant. The data are grouped in two subsets. (b) Data rescaled adding a correction factor $c$. Inset: value of the correction factor $c$ as a function of oil charge relaxation time, showing the two regimes $c \sim$ 1 and $c \sim$ 5, function of $\tau$.}
\label{fig:rescaling}
\end{figure}

It is clear that our simple model cannot be used to predict the exact value of the field across the droplet-droplet since the gap is unknown: determining the stability condition for the film would require to compute the energy density in the film as a function of spacing distance and add it to the energy of interaction of the droplet-oil-droplet system in the absence of field which is also unknown in our case. But they provide a guide to rescale the experimental data. We rescaled the frequencies by the cuttoff frequency of the PDMS/oil RC circuit (\textcolor{red}{Fig. \ref{fig:rescaling}a)}.\\

Interestingly, the data partly collapse and two regimes are observed at high conductivity and low conductivity. This effect was captured in the simulations: we improve the rescaling of the data by a correction factor $c$ and plot $c$ as a function of the charge relaxation time-scale of the oil $\tau = \rho \varepsilon_0 \varepsilon $ (\textcolor{red}{Fig. \ref{fig:rescaling}b}). The correction $c$ is shown to be small ($c \sim 1$) at $\tau < 10^{-4}$ s and constant above ($c \sim 5$), indicating that the PDMS is not the sole responsible for the capacitance in the system and that the dielectric properties of the oil eventually have to be accounted for and dominate at low conductivity.  \\

Our results are compatible with Thiam \textit{et al.} \cite{Thiam2009} showing a minor effect of salt concentration on the coalescence diagram for droplet pairs in flow at small separation distances, and Priest \textit{et al.} \cite{Priest2006} showing a minor effect in static conditions. Szymborski \textit{et al.} observed a weak dependence of the critical voltage for coalescence with salt concentration, with typical variations of the critical voltage in the range 100 - 350 V. However, it is striking to observe that the critical voltage to coalesce the emulsion is of order 100 V over 25 mm when in our case, corresponding to a field of order 100 fold smaller than in our case. The presence of multiple droplets in the coalescence chamber leads to unpredictable field distribution, thereby making the quantitative comparison irrelevant. \\

\begin{figure*}[!h]
\centering
\includegraphics[width=\linewidth]{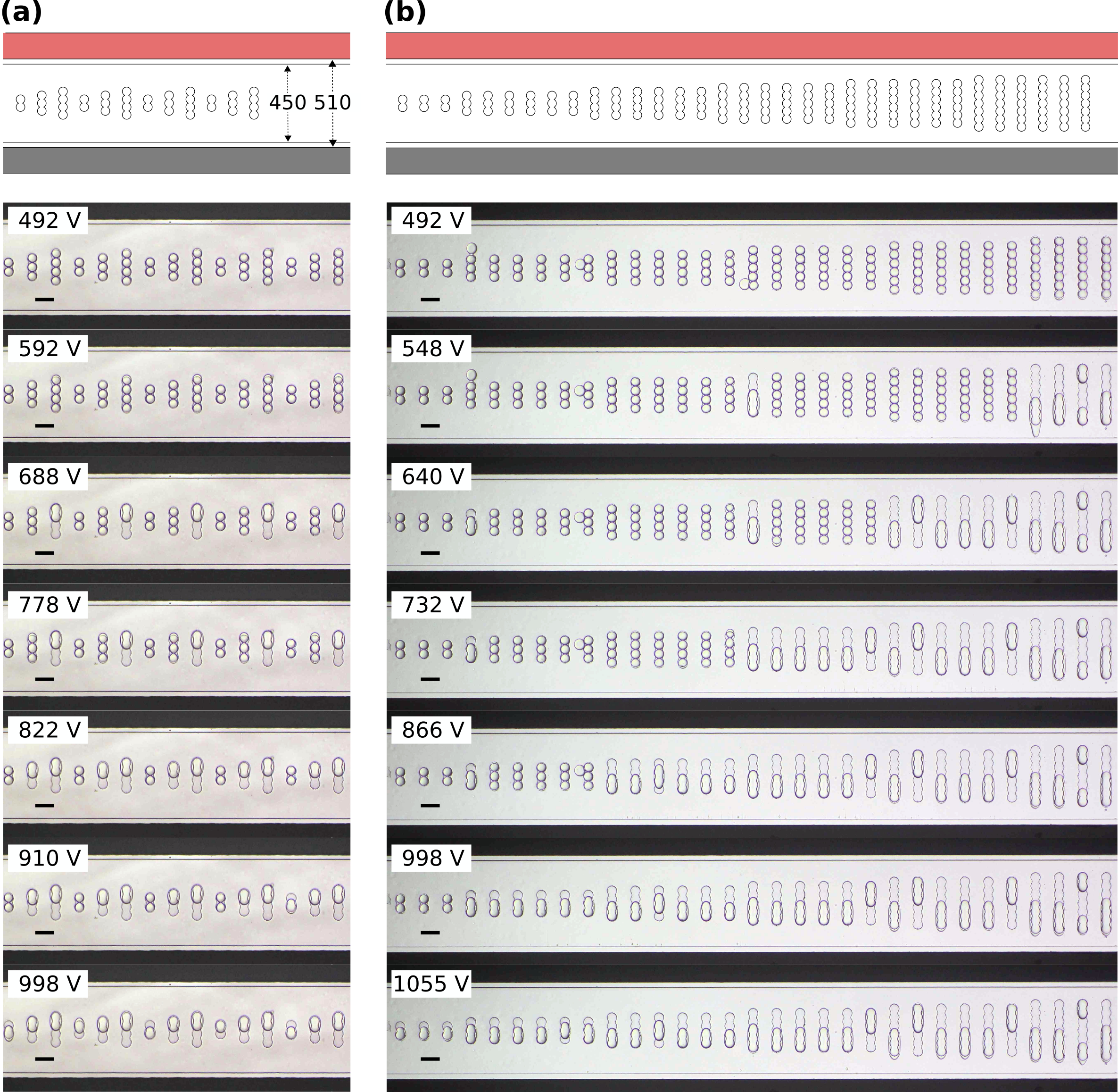}
\caption{Minimal emulsions. Raw pictures showing the impact of the number of droplets on the coalescence voltage on two designs allowing for the analysis up to quadruplets (a) or septuplets (b). Coalescence of the pairs, triplets, quadruplets, up to septuplets induced by increased voltages. Experiments are performed with HFE 7500 at 1\% surfactant. Voltages are applied at a frequency of 10 kHz. }
\label{fig:S8} 
\end{figure*}

\begin{figure*}[!ht]
\centering
\includegraphics*[width=\textwidth]{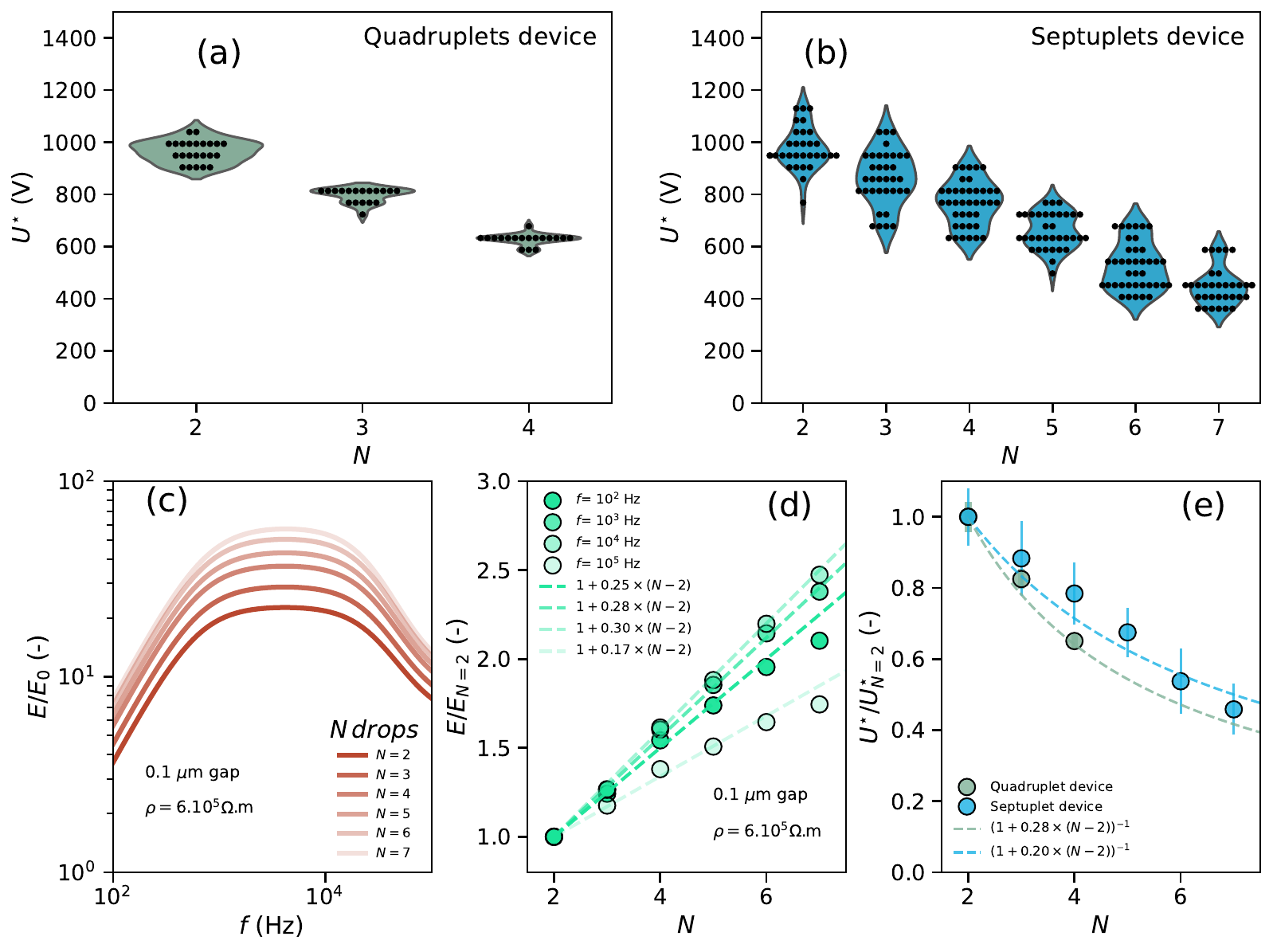}
\caption{From single pairs to emulsion electrocoalescence: lines of droplets in a minimal emulsion are destabilized for voltages smaller than for droplets pairs as a result of local field enhancement. (a) Droplet coalescence for quadruplets analysis. (b) Droplet coalescence for septuplets device. (c) Numerical simulations show that the field is enhanced in emulsions: frequency dependence of the field enhancement. (d) The field enhancement leads to lower threshold voltages for electrocoalescence, linearly dependent on the number of droplets as a first order approximation. (e) The comparison of the field enhancement obtained in the experiments with the linear relationship obtained from the simulation shows a good agreement. }
\label{fig:minemul}
\end{figure*}

In order to make this link between the coalescence of droplet pairs and the coalescence in emulsion, we constructed a minimal emulsion~\cite{Gruner2016}. We experimentally vary the number of droplets in a line along the direction of the field in a channel enlarged to 450 µm width with a spacing distance $d$ of the electrodes of 510 µm (\textcolor{red}{Fig.~\ref{fig:S8}}). We first tested the coalescence of droplet pairs, triplets and quadruplets. The threshold voltage for coalescence decreases from $\sim$ 1 kV for droplet pairs to $\sim$ 0.6 kV for quadruplets at 10 kHz(\textcolor{red}{Fig.~\ref{fig:S8}, Fig.~\ref{fig:minemul}a}). \\

The behaviour is robust: we repeated the experiments on a device designed to generate lines of up to seven droplets. The threshold voltage further decreases down to values smaller than $\sim$ 0.4 kV for septuplets (\textcolor{red}{Fig.~\ref{fig:minemul}b}). It should also be noted that the data become more noisy in this case possibly due to lateral electrical interactions among the droplet chains. \\

We further used our numerical simulations to compute the field between droplets for $N=$ 2 to 7 droplets over the whole frequency range (\textcolor{red}{Fig.~\ref{fig:minemul}c}), showing a systematic enhancement of the field with the number of droplets. For a fixed frequency, the enhancement of the field with $N$ is close to linear with the number of interfaces: $E = E_{N=2}( 1 + \alpha (N-2) )$ where $\alpha$ is a frequency dependent parameter in the range 0.15-0.30 (Fig. ~\ref{fig:minemul}d). For $f=$ 10 kHz, $\alpha$ = 0.28. Knowing the coalescence threshold  for a droplet pair, the threshold for $N$ droplets follows the inverse relationship: $U^{\star}(N) = U^{\star}_{N=2} / ( 1 + \alpha (N-2) )$. This analysis provides means to compare experiments and simulations: for all the experiments, a rescaling of the coalescence voltage by the threshold of the pairs shows that the data follow the same trend. The comparison with the experimental data shows an excellent agreement considering the crude two dimensional assumptions made in the model (\textcolor{red}{Fig. ~\ref{fig:minemul}e}) and the experimental unknowns such as the exact value of the droplet spacing. These experiments nevertheless clearly quantify the coalescence threshold in minimal emulsions, demonstrating the impact of the emulsion geometry on electrocoalescence, related to field enhancement in the oil film separating the droplets. The criterion of coalescence of a pair ($N=2$) therefore controls the coalescence in more complex systems of $N$ droplets with a significant contribution originating from the droplet environment. \\

To summarize our findings, we have shown that the electric parameters of the formulation used, especially the resistivity of the oil controls the coalescence efficiency by changing the cutoff frequency of the equivalent electrical circuit. The decay of the field in the film is the primary source of reduction of the electrocoalescence efficiency, mainly at low field frequency where the threshold can vary over decades. The high frequency limit where the oil can be reduced to a perfect dielectric is experimentally obtained only when the oil conductivity is large enough. For formulations used in practice, this regime might not be observed at all (e.g. with HFE7500-based formulations). In this case, the formulation must be modelled as a leaky dielectric. We then characterized how the threshold of a pair is translated to the coalescence threshold in an emulsion at fixed formulation conditions and show that the field enhancement related to the presence of conducting droplets favors coalescence in emulsions. Our results therefore clarify the scattered reports in the literature of the coalescence threshold and highlights the key parameters affecting the performance of the electrocoalescence process in quiescent conditions.  \\

\pagebreak
\clearpage

\section*{Material and Methods} 

\subsection*{Microfabrication} 
Microfluidic devices are fabricated in polydimethylsiloxane (PDMS) by standard soft lithography methods and replica molding~\cite{Duffy1998}. Electrodes are manufactured symmetrically at both sides of the trapping array by inserting low temperature melting solder (Indium corp.) into the electrode channels, using the microsolidics methods~\cite{Siegel2007}. 

\subsection*{Microfluidic droplet manipulation}
A monodisperse emulsion of aqueous 50~µm droplets is produced with a flow focusing junction in a perfluorinated oil (HFE 7500 or FC 40, 3M) containing a non-ionic surfactant (Fluosurf, Emulseo). The collected emulsion is reinjected in a second microfluidic chip for droplet merging analysis (\textcolor{red}{Supp. Fig. S1}). The second chip consists of a reinjection nozzle with an array of 36 droplet traps. The reinjection module has a height of 27~µm while the droplet traps have a diameter of 50~µm and a height of 60~µm. The flow is controlled with pressurized air (0-200~mbar, Fluigent) and the devices are connected to flow controllers using PTFE tubing with an inner diameter of 0.3~mm (Fischer Scientific). The microfluidic chip is placed on the stage of an inverted microscope (IX71, Olympus) for imaging with a high-speed camera (v210, Phantom). Droplets are loaded in the traps by decreasing the pressure of the most right inlet. Remaining droplets which are not trapped after all the traps are loaded are removed by again increasing the pressure of the right inlet. After removal of the excess droplets the pressure of all inlets is decreased near zero to avoid oil flow in the merging chamber.

\subsection*{Electric actuation} 
One electrode in the device is grounded while an ac voltage is applied to the other electrode by a signal generator (33210A, Agilent) and a high-voltage amplifier which amplified the signal about 1000 times (623B, Trek). The output voltage of the amplifier is calibrated in function of the frequency by an oscilloscope measurement (TDS 2002C, Tektronix) with a high voltage electrode (P5100, Tektronix) (\textcolor{red}{Supp. Fig. S2}). The amplitude of the voltage is manually stepwise increased with 20~V every 5~s until all droplet pairs merge and an image of each setting is recorded (v210, Phantom). The applied voltage is limited to 1000~V, due to the fact that the signal generator had an internal switch in circuit when passing from 990~mV to 1~V. This internal switch causes droplet merging in some cases, therefore not allowing to perform correct interpretation of the experiment beyond this voltage. The fraction of merged droplets is measured for each applied voltage at different frequencies of the field in the range 0 - 50 kHz and for varying formulations (\textcolor{red}{Supp. Fig. S3}). All experiments are reproduced in triplicate ($N=$ 3), to define the error bars in the reported data.

\subsection*{Electric characterization}
The electric properties of the oil phase are characterized by impedance spectroscopy (Impedance Analyzer 7260, Materials Mates). The impedance analyzer is connected to a fluid cell which contains two opposed platinum electrodes (10 mm x 10 mm) spaced at a distance of 10 mm. The cell is filled with the formulation and a frequency scan is performed with 5~V applied. The measurements are corrected with an open and closed circuit measurement to account for the impedance of the equipment. The resisitivity of the aqueous phase is determined with a conductivity meter (CDM 210, Meterlab). The data are summarized in \textcolor{red}{Supp. Table~2}. 

\subsection*{Interfacial tension}
The surface tension was measured with a pendant drop tensiometer (Teclis Scientific) (\textcolor{red}{Supp. Fig.~S4}). A glass cuvette was filled with oil and surfactant while a syringe was filled with millipore water. A reversed needle with an inner diameter of 0.4 mm (100 Sterican, Braun) was used to create a rising water-in-oil droplet. The droplet size ranged from 0.2  to 6~µl. The volume was chosen such that the pendant drop reaches an equilibrium state without detaching. The densities used for calculation of the surface tension are $\rho_{\text{FC 40}} = 1.85$ g/ml, $\rho_{\text{HFE 7500}} = 1.641$ g/ml, $\rho_{\text{water}} = 0.998$ g/ml. The values of the surface tension of the pure water-oil interface were measured to be 46~mN/m for HFE 7500 and 49~mN/m for FC 40. The timescale of the measurement was $>$ 5000 s to reach an equilibrium state. The interfacial tension measured for oil with surfactant were at the lower limit of the method ($\sim$ 5~mN/m) and we assume an accuracy of $\pm$ 0.5 mN/m on the measurement \cite{Brosseau2014}. The data are summarized in \textcolor{red}{Supp. Table~2}. 

\subsection*{Simulations}

The numerical simulations are carried out by resorting to the AC/DC module of the finite-element based commercial software COMSOL Multiphysics. A 2D model, sketched on \textcolor{red}{Supp. Fig. S5(a)} of the problem is built consisting of a square domain of length $L$, bounded vertically by two pads of thickness $t$ and dielectric constant $\varepsilon_\mathrm{PDMS}$ reprensenting the PDMS layers. The coordinate origin is set at the center of the quare domain. Two circles representing the water droplets are positioned at $y = \pm(a+d)/2$, where $a$ is the radius of the circle and $d$ is the seperation distance (gap). The reduced dielectric constant and conductivity of the of the host medium containing the droplets are $\varepsilon$ and $\sigma = 1/\rho$ respectively, while those of the droplets are $\hat{\varepsilon}$ and $\hat{\sigma} = 1/\hat{\rho}$. The top edge of the domain is modeled as an electrode with an electric potential $V$, while the bottom edge is set to a null potential. Periodic boundary conditions are used for the left and right edges. The meshing consists in a fine mesh of free triangles (see \textcolor{red}{Supp. Fig. S5(b)} for a typical mesh realization). A square area is defined around the gap between the two droplets to achieve a finer mesh and correctly resolve the local fields. The maximum elment size within the square is set to $g/10$ (see \textcolor{red}{Supp. Fig. S5(c)}). Additional details on the simulation are provided in the Supporting Method section.

\section*{Acknowledgements}
This project has received funding from the European Union’s Horizon 2020 research and innovation program under the Marie Skłodowska-Curie grant agreement No 813786. JCB aknowledges the support of the `\textit{Fondation Simone et Cino Del Duca}' and of the `\textit{Région Nouvelle Aquitaine}' for funding. We thank Junjin Che (NTG group at CRPP) for his help with the electric characterizations of the fluids used. 

\section*{Conflict of Interest statement}
JCB is cofounder and shareholder of emulseo, producer of the surfactant used in this study.

\bibliographystyle{apsrev4-2}
\bibliography{export}

\end{document}


\title{Emulsion Electrocoalescence in microfluidics: impact of local electric fields}

\author{David Van Assche}
\affiliation{Universit{\'e} de Bordeaux, CNRS, CRPP\unskip, UMR5031\unskip, 115 Avenue Albert Schweitzer\unskip, Pessac\unskip, 33600\unskip, France}

\author{Thomas Beneyton}
\affiliation{Universit{\'e} de Bordeaux, CNRS, CRPP\unskip, UMR5031\unskip, 115 Avenue Albert Schweitzer\unskip, Pessac\unskip, 33600\unskip, France}

\author{Alexandre Baron}
\affiliation{Universit{\'e} de Bordeaux, CNRS, CRPP\unskip, UMR5031\unskip, 115 Avenue Albert Schweitzer\unskip, Pessac\unskip, 33600\unskip, France}
\affiliation{Institut Universitaire de France, 75005 Paris, France}

\author{Jean-Christophe Baret}
\email{jean-christophe.baret@u-bordeaux.fr}
\affiliation{Universit{\'e} de Bordeaux, CNRS, CRPP\unskip, UMR5031\unskip, 115 Avenue Albert Schweitzer\unskip, Pessac\unskip, 33600\unskip, France}
\affiliation{Institut Universitaire de France, 75005 Paris, France}

\date{\today}

\maketitle

\section*{Supporting Methods}

\subsection*{Simulations}

\begin{figure}[t!]
	\centering	
    \includegraphics[width=\textwidth]{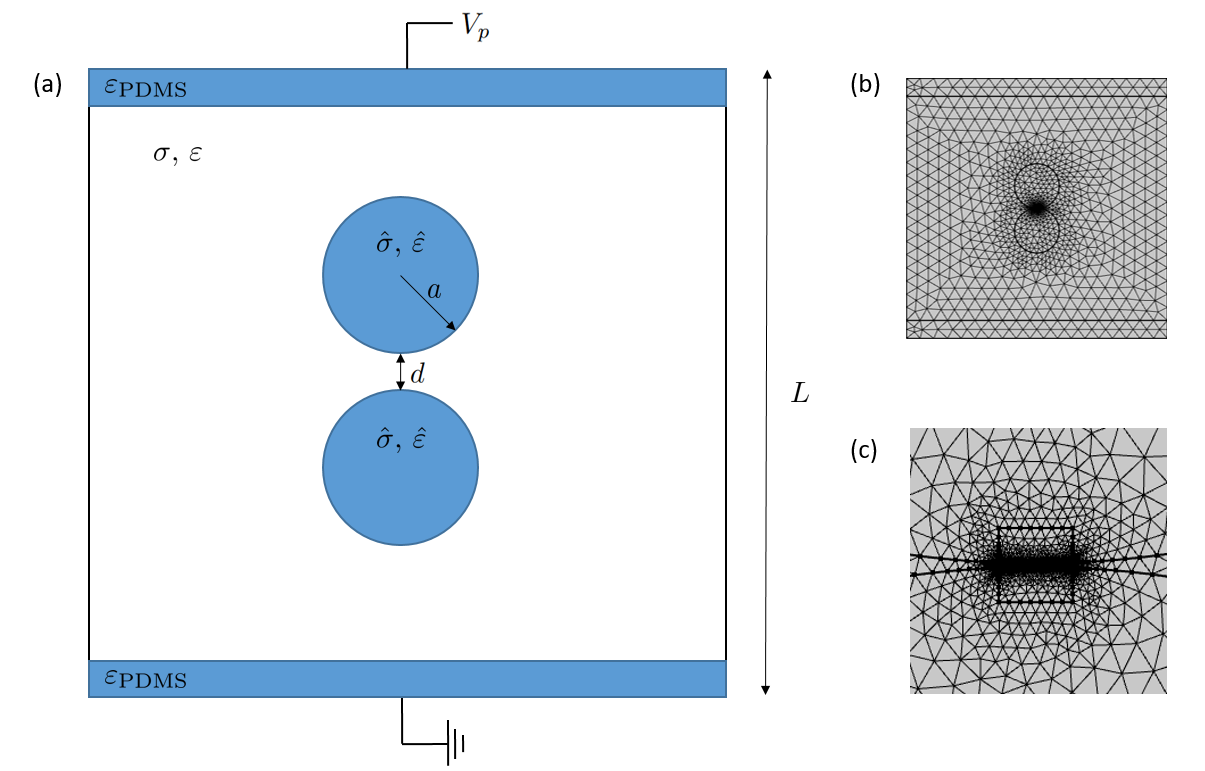}
	\caption{\label{Fig:stack} Numerical finite-element model. (a) Sketch of the 2D model used to simulate the experimental situation with parameters shown. (b) Typical mesh of the overall simulation domain. (c) Refined mesh in a square domain near the gap area.}
	\label{Fig:2D_model}
\end{figure}

The numerical simulations are carried out by resorting to the AC/DC module of the finite-element based commercial software COMSOL Multiphysics. A 2D model, sketched on Fig. \ref{Fig:2D_model}(a) of the problem is built consisting of a square domain of length $L$, bounded vertically by two pads of thickness $t$ and dielectric constant $\varepsilon_\mathrm{PDMS}$ reprensenting the PDMS layers. The coordinate origin is set at the center of the quare domain. Two circles representing the water droplets are positioned at $y = \pm(a+d)/2$, where $a$ is the radius of the circle and $d$ is the seperation distance (gap). The reduced dielectric constant and conductivity of the of the host medium containing the droplets are $\varepsilon$ and $\sigma = 1/\rho$ respectively, while those of the droplets are $\hat{\varepsilon}$ and $\hat{\sigma} = 1/\hat{\rho}$. The top edge of the domain is modeled as an electrode with an electric potential $V$, while the bottom edge is set to a null potential. Periodic boundary conditions are used for the left and right edges. The meshing consists in a fine mesh of free triangles (see Fig. \ref{Fig:2D_model}(b) for a typical mesh realization). A square area is defined around the gap between the two droplets to achieve a finer mesh and correctly resolve the local fields. The maximum elment size within the square is set to $g/10$ (see Fig. \ref{Fig:2D_model}(c)).

The software solves the electrostatic problem defined as: 
\begin{eqnarray}
    \nabla.\D &=& 0\\
    \E &=& -\nabla V
\end{eqnarray}
where $\D = \varepsilon_0 (\varepsilon-i\frac{\sigma}{\varepsilon_0\omega})\E$ is the electric displacement field, $\E$ is the electric field and $V$ is the electric potential.

In the domain surrounding the droplets (oil phase), the maximum value of the electric field is retrieved and normalized to $E_0 = V/L$. This value is computed as a function of frequency for varying gap size $g$ and conductivity~$\sigma$.

\begin{figure*}[!h]
\centering
\includegraphics[width=0.8\linewidth]{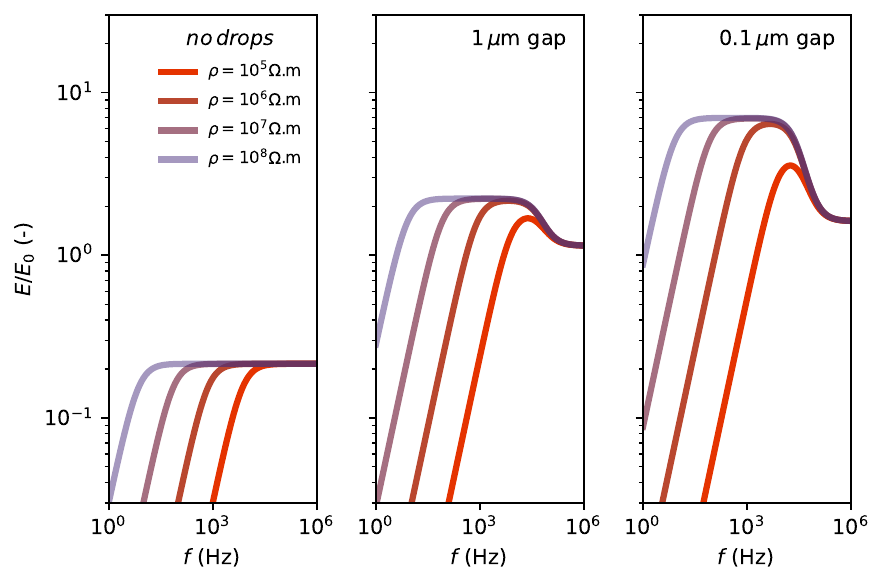}
\includegraphics[width=0.8\linewidth]{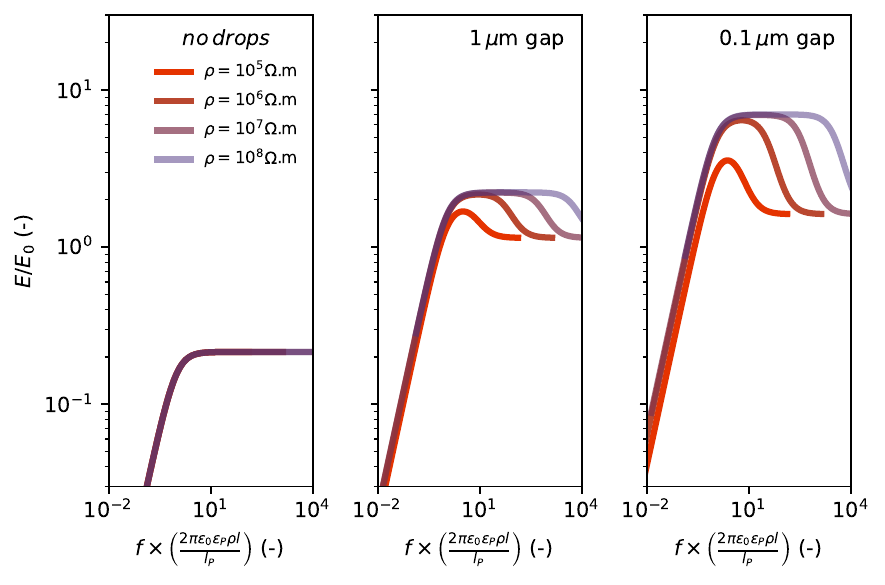}
\caption{Numerical simulations of the maximum field enhancement $E/E_0$ within the domain surrounding the droplets (oil) as a function of frequency (top three panels) and as a function of reduced frequency for various values of the resistivity $\rho = 1/\sigma$.}
\label{fig:S7} 
\end{figure*}

In the absence of droplets, the field is shown to depend on the frequency as a high pass filter with a cutoff given by the RC constant of the PDMS/oil system (Supp. Figure~\ref{fig:S7}). In the presence of the droplet, field enhancement is observed, depending both on the droplet spacing and frequency. These simulations explain why the electrocoalescence mechanism strongly depends on frequency. For this simulation the parameters are set as follows: $a=25$ $\mu$m, $L=250$ $\mu$m, $\hat{\sigma}=370$ $\mu$S/m, $\varepsilon=9.8$ and $\hat{\varepsilon}=80$.

\begin{figure*}[!h]
\centering
\includegraphics[width=0.8\linewidth]{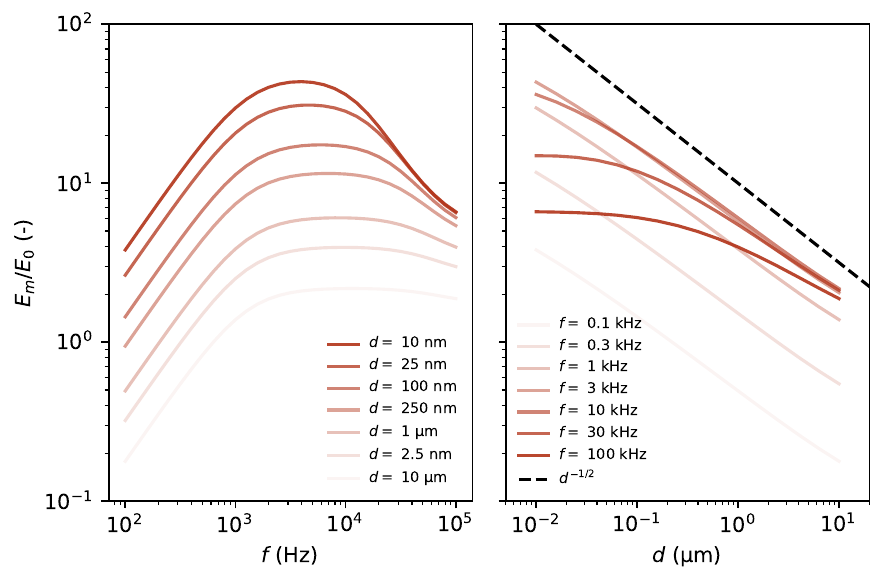}
\caption{Numerical simulations of the maximum field enhancement $E/E_0$ within the domain surrounding the droplets (oil) as a function of frequency for various gap sizes (left panel) and as a function of gap size for various frequencies (right panel).}
\label{fig:S7-2} 
\end{figure*}

The dependence of the electric field with frequency for various gaps is plotted on Supp. Fig.\ref{fig:S7}, \ref{fig:S7-2}. Interestingly, the experiments show that the field is enhanced with a strong resonance in the kHz regime, especially at small gap dimensions. The dependence of the field with the gap is non trivial with a -1/2 exponent which implies that the voltage across the film does not linearly depend on the droplet spacing.  

For the simulations carried out to produce the data of Supp. Fig.~\ref{fig:S7},\ref{fig:S7-2},  we set $L=600$ and $\sigma= 1.6$ $\mu$S/m. The rest of the parameters are kept the same as previously. These data support the absence of universal scaling for the electrocoalescence mechanism. 

\subsection*{Electrical model}

Qualitatively, the numerical simulations are approximated using a simple equivalent electrical circuit. To model the dependence of $U^*_{applied}$ as a function of $f$, we propose a minimal equivalent electric circuit that allows to estimate the voltage across the oil film as a function of the system parameters (Supp. Figure~\ref{fig:model}). We use a simple voltage divider model using resistors and capacitors where the impedances are given by the fluid properties and the geometry of the system according to Equation \ref{model}: 
\begin{equation}
   \frac{U_{film}}{U_{applied}} = \left(1 + 2 \frac{Z_{drop} + Z_{pdms} + Z_{oil}}{Z_{film}} \right)^{-1}
  \label{model}
\end{equation}

The impedances $Z$ of electrical components are modelled as a capacitor and resistor in parallel, which results in the impedance: 
$$Z_{RC} = \frac{R}{1+i\omega CR}$$
The PDMS layer is considered a perfect capacitor with capacitance $C$: $Z_C = 1/i\omega C$ with $\omega = 2\pi f$. The drops are modeled as a pure resistor $R_{drop}$. The elements are modelled in a planar approximation, with $\rho_{water}$ and $\rho_{oil}$ the resistivity of the water and the oil phase (in $\Omega \cdot m$), $\varepsilon_0$ the vacuum permittivity of $8.85\cdot 10^{-12}F/m$, $\varepsilon_{r,pdms}$, $\varepsilon_{r,oil}$ the relative permittivity of the PDMS and the oil phase. $l$, $h$ and $w$ are the geometrical parameters specific to the microfluidic chip design. The geometrical dimensions used to solve the model are presented in Table \ref{table:properties}. The model data are present in Supp. Figure~\ref{fig:model}, showing the impact of the oil resistivity on the voltage across the film.

\subsection*{Minimal emulsion electrocoalescence}

Devices to study the minimal emulsions are manufactured as described above. The experiments are analyzed according to the same protocol. The threshold voltage for coalescence is determined for each configuration from pairs to septuplet in two different devices.   

\pagebreak
\clearpage

\section*{Supporting Figures and Supporting Tables}

\begin{figure*}[!ht]
\centering
\includegraphics*[width=0.7\linewidth]{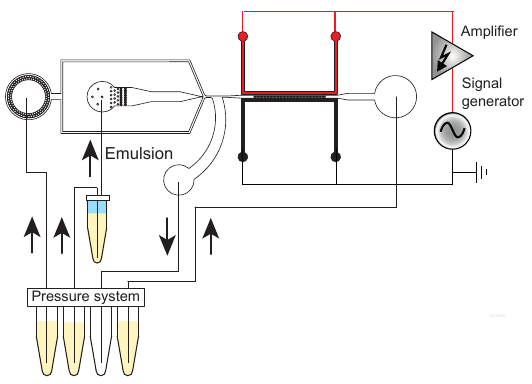}
\caption{Microfluidic device for reinjection and trapping of 50 µm droplets. Droplets are reinjected and loaded in the traps by manipulating the pressure at the inlets.} \label{fig:S2}
\end{figure*}

\pagebreak
\clearpage

\begin{figure*}[!ht]
\centering
\includegraphics*[width=0.7\linewidth]{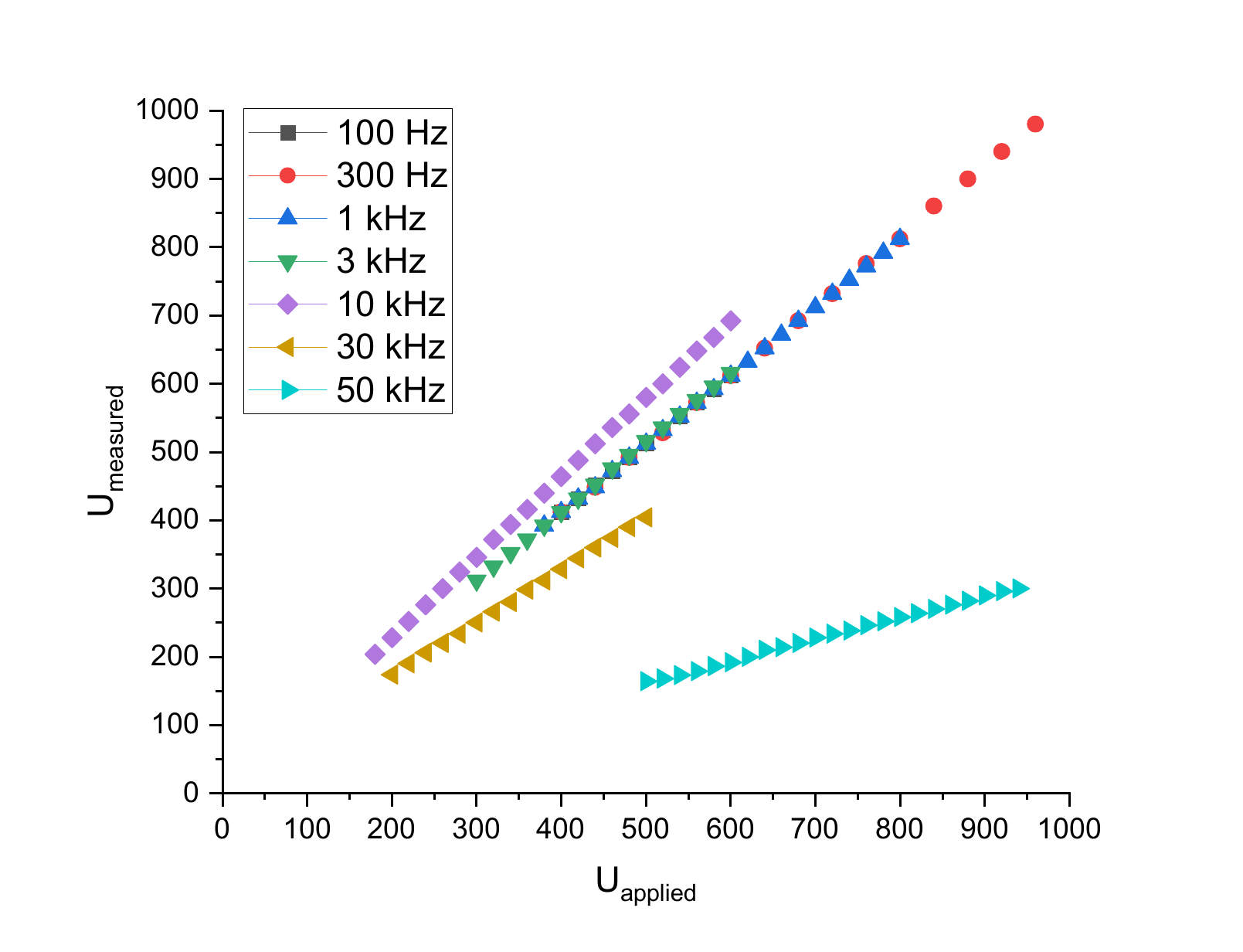}
\caption{Calibration for output voltage of the measured high voltage amplifier (623B, Trek) as a function of the applied voltage (Agilent low frequency generator).} \label{fig:S1}
\end{figure*}

\pagebreak
\clearpage

\begin{figure*}[!h]
\centering
\includegraphics[width=1\linewidth]{raw data.pdf}
\caption{Fraction of merged droplet pairs as a function of $U_{applied}$ for different frequencies of the electric field. (a) HFE 7500 0.1\% w/w surfactant and millipore drops; (b) HFE 7500 1\% w/w surfactant and millipore drops; (b) HFE 7500 5\% w/w surfactant and 2 M NaCl drops; (d) FC 40:HFE 7500 (0.5:0.5) 5\% w/w surfactant and millipore drops; (e) FC 40:HFE 7500 (0.75:0.25) 5\% w/w surfactant and millipore drops; (f) FC 40 5\% w/w surfactant and millipore drops; (g) FC 40 5\% w/w surfactant and $4\cdot10^{-4}$ M NaCl drops; (h) FC 40 5\% w/w surfactant and 2 M NaCl drops.}
\label{fig:S3}
\end{figure*}

\pagebreak
\clearpage

\begin{figure*}[!h]
\centering
\includegraphics*[width=0.85\linewidth]{surface tension.pdf}
\caption{Surface tension measurements for (a) mixtures of FC 40 and HFE 7500 at 5\% w/w surfactant concentration, and (b) HFE 7500 at 0.1, 1 and 5\% w/w surfactant concentration.} \label{fig:S4}
\end{figure*}

\pagebreak
\clearpage

\pagebreak
\clearpage

\pagebreak
\clearpage

\pagebreak
\clearpage

\begin{figure*}[!h]
\centering
\includegraphics[width=0.8\linewidth]{fig electricmodel.pdf}
\caption{Minimal electrical model. (a) Equivalent electrical circuit of one trapped droplet pair. The oil is modeled as a capacitor and resistor in parallel. (b) Calculation of $U_{film}/U_{applied}$ when putting in the experimentally measured parameters and geometrical estimations. (c) $U_{film}/U_{applied}$ rescaled with $f_{0, model}$. The data collapses for $f_{0, model} \sim \rho^{-1}_e$.}
\label{fig:model} 
\end{figure*}

\newpage 
\clearpage

\begin{table*}[!h]
\caption{}
\begin{ruledtabular}
\begin{tabular}{p{0.7cm}p{3.5cm}p{5cm}p{3cm}}
 Element & Impedance & equivalent $R$, $C$& Geometry values \\
 \hline
 PDMS        &   $Z_{pdms}  = 1/i\omega C_{pdms} $  & $C_{pdms} = \varepsilon_0 \varepsilon_{r, pdms}\frac{l_{pdms}h_{pdms}}{w_{pdms}}$ &     $h_{pdms} = 50~\mu$m \newline  $l_{pdms} = 50~\mu$m \newline $w_{pdms} = 20~\mu$m\\
 
 Oil (bulk)  &   $Z_{oil} = \frac{R_{oil}}{1+i\omega C_{oil}R_{oil}}$ & $R_{oil} = \rho_{oil} \frac{w_{oil}}{l_{oil}h_{oil}}$ \newline $C_{oil} = \varepsilon_0 \varepsilon_{r, oil}\frac{l_{oil}h_{oil}}{w_{oil}}$  &   $h_{oil} = 50~\mu$m  \newline $l_{oil} = 50~\mu$m  \newline  $w_{oil} = 20~\mu$m\\

 Oil (film)  &   $Z_{film} = \frac{R_{film}}{1+i\omega C_{film}R_{film}}$ &    $R_{film} = \rho_{oil} \frac{w_{film}}{l_{film}h_{film}}$  \newline $C_{film} = \varepsilon_0 \varepsilon_{r, oil}\frac{l_{film}h_{film}}{w_{film}}$  &     $h_{film}= 10~\mu$ m  \newline  $l_{film} = 10~\mu$m  \newline $w_{film} = 1~\mu$m \\
 Droplet     &   $Z_{drop} = R_{drop}$ &   $R_{drop} = \rho_{water} \frac{w_{drop}}{l_{drop}h_{drop}}$ &   $w_{drop} = 50~\mu $m \newline  $l_{drop} = 50~\mu$m \newline  $h_{drop} = 50~\mu$m\\

 \end{tabular}
\end{ruledtabular}\label{table:properties}
\end{table*}

\newpage 
\clearpage

\begin{table*}[!h]
\caption{Material properties of the fluids used in the experiments.}
\begin{ruledtabular}
\begin{tabular}{p{2.5cm}p{2.5cm}p{2.5cm}p{2.5cm}p{2.5cm}}
 Fluid & Surfactant concentration & Resistivity $\rho$ & Dielectric constant $\varepsilon$ & Surface tension $\gamma$\\ 
 & \% w/w & $10^5~\Omega \cdot$m & & mN/m\\ \hline
 HFE 7500 & 0.1 & $17.0 \pm 1.0$ & $9.62 \pm 0.01$ & 2.7 \\ 
 & 1  & $6.4\pm 0.3$ & 9.62 $\pm$ 0.003& 3.7\\
 & 5 & $1.9\pm 1.0$& 9.76 $\pm$ 0.26 & 3.9 \\
 HFE 7500 - FC40 & 5 & $9.5\pm 1.0$ & 6.63 $\pm$ 0.85 & 7.0 \\
  (0.5:0.5) &  & & & \\
 HFE 7500 - FC40 & 5 & $49.0\pm 4.2$ &4.76 $\pm$ 0.001 & 5.2 \\
  (0.25:0.75) &  & & & \\
 FC40  & 5& $463.0\pm 17.1$ &  3.52 $\pm$ 0.22 & 5.3 \\ \hline \hline
  Fluid & NaCl concentration & Resistivity $\hat{\rho}$ & Dielectric constant $\hat{\varepsilon}$ & \\ 
    & M & $\Omega \cdot$m & & \\ \hline
  Water & 0 & 2710& $78.43 \pm 0.10$ \cite{water_dielectric} &-\\
 &$4\cdot 10^{-4}$  & 117.2 & - &-\\
 &2 & $69.7\cdot 10^{-3}$& -&-\\
\end{tabular}
\end{ruledtabular}\label{table:properties}
\end{table*}

\bibliographystyle{apsrev4-2}
\bibliography{export.bib}